\documentstyle[sprocl]{article}

\def\be{\begin{equation}}
\def\ee{\end{equation}}
\def\bea{\begin{eqnarray}}
\def\eea{\end{eqnarray}}

\newcommand{\tr}{{\,\rm tr\,}}

\newcommand{\bda}{\begin{\displaymath}\begin{array}{rl}}
\newcommand{\eda}{\end{array}\end{displaymath}}
\newcommand{\bdm}{\begin{displaymath}}
\newcommand{\edm}{\end{displaymath}}

\newcommand{\fs}{\; .}
\newcommand{\co}{\; ,}
\newcommand{\eff}{{e\hspace{-0.1em}f\hspace{-0.18em}f}}

\newcommand{\indA}{{\scriptscriptstyle A}}

\newcommand{\indL}{{\scriptscriptstyle L}}
\newcommand{\indR}{{\scriptscriptstyle R}}

\newcommand{\nc}{N_{\!c}}

\newcommand{\bce}{\begin{center}}
\newcommand{\ece}{\end{center}}

\begin{document}

\title{CHIRAL PERTURBATION THEORY AND THE $1/\nc$-EXPANSION}

\author{ROLAND KAISER}

\address{Institute for theoretical Physics, University of Bern,\\ 
Sidlerstr. 5, CH-3012 Bern, Switzerland\\E-mail: kaiser@itp.unibe.ch} 

\maketitle
\abstracts{We briefly review the effective theory that 
  describes the low energy properties of QCD with three light quarks and a
  large number of colours, $\nc$, and then discuss the mechanisms that forbid
  the 
  Kaplan-Manohar  
  transformation in this framework.}

\section{Expansion in the quark masses and $1/\nc$}

It is generally accepted\footnote{See ref. \cite{Kaiser:2000gs} for a detailed
  list 
  of references} that QCD possesses
an exact  
U(3)$_\indR\times$U(3)$_\indL$ flavour symmetry in the limit $m_u \! =\! m_d\!
=\! m_s \!=0 $, $\nc\!\to\!\infty$. We furthermore assume that (a) the axial
  part 
of this symmetry is spontaneously broken and (b) that, in the
vicinity of this point, the low energy properties of the theory are governed by
the Goldstone bosons associated with this symmetry breakdown. These
  assumptions provide the basis for the effective theory, where the nine
  Goldstone degrees of freedom are collected in
a matrix $U(x) \in  $ U(3). 
Apart from the Wess-Zumino term, the effective Lagrangian represents the most
general expression 
formed with the field $U$ and the source fields
for the quark currents that is consistent with chiral symmetry. Furthermore, in order to study the consequences of the
U(1)$_\indA$-anomaly, we include a source field 
for the 
winding number density $\omega=1/(16\pi^2) 
\tr_{\!\!c}\,
G_{\mu\nu}\widetilde{G}^{\mu\nu}$. We denote this field by $\theta(x)$; in the
  QCD-Lagrangian it enters in the form ${\cal L}_{\mbox{\tiny QCD}} =-
  \theta \, 
  \omega +... $  

The terms in the effective Lagrangian are ordered by introducing a counting
parameter $\delta $,  
${\cal L}_\eff = {\cal L}_{1}+{\cal L}_{\delta}+{\cal L}_{\delta^2}+ \ldots $
A coherent expansion
emerges if powers of momenta, quark masses and $1/\nc$ are counted according
to \cite{Leutwyler:1996sa}   
\bea\label{deltacount}
p = O(\sqrt{\delta })\co\;\;\; m  = O (\delta )  \co\;\;\; 1/\nc  = O (\delta
)  \fs 
\eea
The effective coupling constants are proportional to powers of $\nc$. To
determine these, one 
infers the 
known large $\nc$ 
behaviour of the QCD correlation functions~\cite{Gasser:1985gg}. These state
in particular that correlation functions involving the  
winding number density $n$ times are 
suppressed by a factor of $\nc^{-n}$. 
As shown in ref. \cite{Kaiser:2000gs}, this implies that the dependence of the
effective Lagrangian on the field  
$\theta$ is polynomial, to every order in the expansion in
powers of $\delta$. 

\section{The Kaplan-Manohar transformation}

Kaplan and Manohar \cite{Kaplan:1986ru} pointed out the existence of an
ambiguity in the 
quark mass ratios inherent to the three flavour effective
Lagrangian at fixed $\nc$. 
A priori, this situation persists also if the number of colours
is taken large: In the effective Lagrangian, one may replace
the quark mass matrix $m$ with
\bea\label{KMT}
m' =  m+\lambda\, e^{-i\theta}  m^{\dagger-1} \det m^\dagger  \co
\eea
without violating chiral symmetry. Under a chiral transformation, the
mass matrix transforms according to $m 
\to V_\indR m {V_\indL}^{\!\dagger} $, with $ V_\indR,{V_\indL} \in$ U(3). The
matrix $m'$ transforms in the same manner
since the 
factor 
  $e^{- i \theta}$
transforms contragrediently to $\det m^\dagger $ under
U(1)$_\indA$-rotations. The Lagrangian ${\cal L}_\eff (m')$ is, however,
in conflict with large $\nc$, 
because, in view of the factor $\lambda\, e^{- i \theta}$, it involves
arbitrary 
powers 
of 
the field $\theta$. 
As stated in the preceding section, the
original Lagrangian, ${\cal L}_\eff (m)$, depends on this variable only
polynomially. 
The conflict is resolved only 
if the parameter $\lambda$ 
vanishes to all orders in $1/\nc$. \cite{Kaiser:2000gs}

There is no reason why the effective theory should allow a
transformation of the type (\ref{KMT}) in the first place -- after all QCD
possesses  
no 
corresponding 
symmetry. It is for instance  
well-known that correlation functions of 
the scalar or pseudoscalar currents are not invariant under a change of the
mass  
matrix. Experimental information on these would pin
down the values of the 
quark 
mass ratios also in the standard framework \cite{Leutwyler:1990pn}.  
The generalized version of the
Kaplan-Manohar transformation in eq.~(\ref{KMT}) necessarily involves the
vacuum angle $\theta$, as a consequence the 
correlation functions of the winding number density $\omega$ also fail to be
invariant. We do not have experimental information on these either, but they
are constrained theoretically at large $\nc$. As it happens, the
transformation in eq.~(\ref{KMT}) violates these constraints.

\section*{Acknowledgments}
I wish to thank the organizers of this nice workshop and  
acknowledge the support by the Swiss National Science
Foundation. This article represents a report on work done in
collaboration with H. Leutwyler.

\end{document}